\input phyzzx.tex
\tolerance=1000
\voffset=-0.0cm
\hoffset=0.7cm
\sequentialequations
\def\rl{\rightline}

\def\t1{{\tilde 1}}

\def\t{\theta}

\def\S{Schwarzschild~}
\def\s{superconformal~}

\REF{\BEK}{J. Bekenstein, Lett. Nuov. Cimento {\bf 4} (1972) 737; Phys Rev. {\bf D7} (1973) 2333; Phys. Rev. {\bf D9} (1974) 3292.}
\REF{\HAW}{S. Hawking, Nature {\bf 248} (1974) 30; Comm. Math. Phys. {\bf 43} (1975) 199.}
\REF{\MAL}{J. M. Maldacena, hep-th/9607235; G. T. Horowitz, gr-qc/9704072; A. W. Peet, hep-th/0008241.}
\REF{\LEN}{L. Susskind, hep-th/9309145.}
\REF{\SBH}{E. Halyo, A. Rajaraman and L. Susskind, Phys. Lett. {\bf B392} (1997) 319, hep-th/9605112.}
\REF{\HRS}{E. Halyo, B. Kol, A. Rajaraman and L. Susskind, Phys. Lett. {\bf B401} (1997) 15, hep-th/9609075.}
\REF{\EDI}{E. Halyo, Int. Journ. Mod. Phys. {\bf A14} (1999) 3831, hep-th/9610068; Mod. Phys. Lett. {\bf A13} (1998), hep-th/9611175.}
\REF{\VEN}{T. Damour and G. Veneziano, Nucl. Phys. {\bf B568} (2000) 93, hep-th/9907030.}
\REF{\POL}{G. T. Horowitz and J. Polchinski, Phys. Rev. {\bf D55} (1997) 6189, hep-th/9612146; Phys. Rev. {\bf D57} (1998) 2557, hep-th/9707170.}
\REF{\DAM}{T. Damour, gr-qc/0104080.}
\REF{\UNI}{E. Halyo, JHEP {\bf 0112} 005 (2001), hep-th/0108167.}
\REF{\KAU}{R. Kaul, Phys. Rev. {\bf D68} (2003) 024026, hep-th/0302170.}
\REF{\BIT}{E. Halyo, hep-th/0308166.}
\REF{\HST}{G. T. Horowitz and A. Strominger, Nucl. Phys. {\bf B360} (1991) 197.}
\REF{\ULF}{U. Danielsson, A. Guijosa and M. Kruczenski, JHEP {\bf 0109} 011 (2001), hep-th/0106201.}
\REF{\GUI}{A. Guijosa, H. H. Hernandez and H. A. Morales Tecotl, hep-th/0402158.}
\REF{\PEE}{O. Saremi and A. W. Peet, hep-th/0403170.}
\REF{\ORE}{O. Bergman and G. Lifschytz, hep-th/0403189.}
\REF{\RAM}{S. Kalyana Rama, hep-th/0404026.}
\REF{\GIL}{G. Lifschytz, hep-th/0405042.}
\REF{\KRS}{S. Kalyana Rama and S. Siwach, hep-th/0405084.}
\REF{\TOR}{W. Fischler, E. Halyo, A Rajaraman and L. Susskind, Nucl. Phys. {\bf B501} (1997) 409, hep-th/9703102.}
\REF{\SEI}{N. Seiberg, hep-th/9705117.}
\REF{\ITZ}{N. Itzhaki, J. M. Maldacena, J. Sonnenschein and S. Yankielowicz, Phys. Rev {\bf D58} (1998) 046004, hep-th/9802042.}
\REF{\ROZ}{M. Rozali, Phys. Lett. {\bf B400} (1997) 260, hep-th/9702136.}
\REF{\BRS}{M. Berkooz, M. Rozali and N. Seiberg, Phs. Lett. {\bf B408} (1997) 105, hep-th/9704089.}
\REF{\MST}{J. M. Maldacena, Nucl. Phys. {\bf B477} (1996) hep-th/9605016; J. M. Maldacena and A. Strominger, JHEP {\bf 9712} 008 (1997), hep-th/9710014.} 
\REF{\KLE}{I. R. Klebanov and A. A. Tseytlin, Nucl. Phys. {\bf B475} (1996) 164, hep-th/9604089.}
\REF{\MAT}{T. Banks, W. Fischler, S. Shenker and L. Susskind, Phys. Rev. {\bf D55} (1997) 5112, hep-th/9610043.}
\REF{\MBH}{T. Banks, W. Fischler, I. R. Klebanov and L. Susskind, Phys. Rev. Lett. {\bf 80} (1998) 226, hep-th/9709091; JHEP {\bf 9801} (1998) 008, hep-th/9711005.}
\REF{\KLS}{I. R. Klebanov and L. Susskind, Phys. Lett. {\bf B416} (1998) 62, hep-th/9709108.}
\REF{\EDH}{E. Halyo, hep-th/9709225.}
\REF{\HOR}{G. T. Horowitz and E. J. Martinec, Phys. Rev. {\bf D57} (1998)4935, hep-th/9710217.}
\REF{\KAL}{S. Kalyana Rama, Phys. Rev. {\bf D59} (1999) 02400, hep-th/9806225.}
\REF{\DAS}{S. R. Das, S. D. Mathur, S. Kalyana Rama and P. Ramadevi, Nucl. Phys. {\bf B527} (1998) 187, hep-th/9711003.}
\REF{\DES}{E. Halyo, hep-th/0107169.}

\singlespace
\rl{SU-ITP-04-}
\rl{hep-ph/0406082}
\rl{\today}
\pagenumber=0
\normalspace
\medskip
\bigskip
\titlestyle{\bf{Black Hole Entropy and Superconformal Field Theories on Brane-Antibrane Systems}}
\smallskip
\author{ Edi Halyo{\footnote*{e--mail address: halyo@itp.stanford.edu}}}
\smallskip
\centerline{Department of Physics}
\centerline{Stanford University}
\centerline{Stanford, CA 94305}
\smallskip
\vskip 2 cm
\titlestyle{\bf ABSTRACT}

We obtain the entropy of \S and charged black holes in $D>4$ from \s gases that live on $p=10-D$ dimensional brane--antibrane systems wrapped on $T^p$. The properties of 
strongly coupled \s theories such as the appearance of hidden dimensions (for $p=1,4$) and fractional strings (for $p=5$) are crucial for our results. In all cases, the \S radius is  
given by the transverse fluctuations of the branes and antibranes due to the finite temperature. We show that our results can be generalized to multicharged black holes.

\singlespace
\vskip 0.5cm
\endpage
\normalspace

\centerline{\bf 1. Introduction}
\medskip

The origin of black hole entropy[\BEK,\HAW] is one of the main open questions in string theory. The entropy of extreme and near--extreme black holes can be successfully
calculated in string theory[\MAL]. However, the entropy of \S black holes still remains unexplained. \S black holes can be described as very highly excited strings at the Hagedorn temperature
and living on the stretched horizon[\LEN,-\BIT]. For asymptotic observers, these strings
have rescaled tensions due to the gravitational redshift. This description gives the entropy of \S black holes and black $p$--branes[\HST] for any amount of nonextremality and in all dimensions 
up to a numerical factor of $O(1)$.
Due to the complete lack of supersymmetry and our lack of knowledge of the physics of very highly excited strings, it is quite difficult to improve this description.

Recently, a different description of \S black holes was given in ref. [\ULF] in terms of a brane--antibrane system. It was shown that the entropy of $D=7$ \S 
black holes can be obtained from two copies of a $3+1$ dimensional Yang--Mills gas at a finite temperature living on an equal number of $D3$ and ${\bar D}3$ branes which are wrapped on a $T^3$. 
The branes and antibranes do not annihilate each other and the system is stable due to the finite (and equal) temperatures of the gases which result in a positive mass squared to the tachyon.  
The prescription of ref. [\ULF] is, for a given mass and charge,
to maximize the entropy of the gas. This fixes the number of branes and the energy of the gas (nonextreme energy) which is of the same order as the brane tension.
This prescrition gives the correct entropy for \S black holes in $D=7$ dimensions in terms of nondilatonic branes, i.e. $D3$ branes of string theory.
The prescription also gives the entropy of $D=7$ charged black holes (i.e. wrapped black branes) simply by taking different numbers of branes and antibranes. 
The work of ref. [\ULF] has been
generalized to cases in other dimensions by using near--extreme black brane entropy formulas for unstable branes or $D {\bar D}$ systems far from extremality[\PEE,\ORE]. Other 
generalizations such as rotating and/or multi--charged black holes appeared in refs. [\GUI,\RAM,\GIL,\KRS].

In this paper we show that the entropy of \S and charged black holes in $D>4$ can be explained by that of a pair of superconformal gases at a finite temperature living on
a system of $N$ branes and $\bar N=N$ antibranes of dimension $p=10-D$. We argue that, in general, for a given $p$ brane (or antibrane) the world--volume theory is not the $p+1$ dimensional 
ideal Yang--Mills gas but the corresponding superconformal gas due 
to the very large dimensionless 't Hooft coupling. Due to our lack of knowledge of the precise numerical coefficients in the equations of state for these \s theories, our results agree with
the black hole entropy up to factors of $O(1)$. However, if these coefficients are obtained from the AdS/CFT duality, the mismatch is a calculable power of $2$[\PEE,\GIL,\KRS]. 
For $p=3$, $D=3+1$ ${\cal N}=4$ supersymmetric Yang--Mills theory is superconformal for any value of the coupling which agrees 
with ref. [\ULF]. (Similarly the prescription of ref. [] applies to $M2$ and $M5$ branes because the world--volume theories of these are also \s.) Interesting but well--known properties of 
$p+1$ 
dimensisonal \s field theories[\TOR,\SEI,\ITZ] are crucial for our description. For example, in the \s limit, the $1+1$ dimensional theory grows a new dimension of size $1/g$ whereas 
a $4+1$ dimensional
theory grows a dimension of size $g^2$[\ROZ,\BRS]. In $5+1$ dimensions, the world--volume theory is described by fractional strings[\MST]. We consider these cases in detail and then 
give the general equation of state for the \s gas. We use the 
prescription of ref. [\ULF] to find the correct entropy formula (up to numerical factors of $O(1)$) for the corresponding \S black holes. The temperature of the \s gas
gives the Hawking temperature in all cases. We show that charged black hole entropy in $D>4$ can be obtained by taking $N \not= \bar N$ where the charge is $Q=N- \bar N$. In this case the \s
on the branes and antibranes have different temperatures. For \S and charged black holes (in any $D>4$) we show that the \S radius is given by the extent of the transverse thermal fluctuations
of the branes and antibranes.
We also generalize these results to multicharged black holes. These are described by $2^n$ different types of \s gas (each at a different temperature)
which arise from the different types of strings that connect the different branes and/or antibranes. Since the total entropy is given by the sum of the entropies of the \s gases,
it seems that these \s gases do not interact with each other.

This paper is organized as follows. In section 2, we briefly review the presciption of ref. [\ULF]. In section 3, we consider \S black holes in $D>4$. We the examine the cases of $D=5,6,9$ in 
detail and then give general results for \S black holes in $D>4$. In section 4, we consider black $p$ branes, i.e. charged black holes which may or may not be near extreme. In section 5, w
e generalize our results to multicharged black holes. Section 6 contains a discussion of our results and our conclusions.

\bigskip
\centerline{\bf 2. $D=7$ Black Hole Entropy and Brane--Antibrane Systems}

In ref. [\ULF], the entropy of $D=7$ \S black holes were obtained from a gas living on $N$ $D3$--${\bar D}3$ branes wrapped on $T^3$. In this setup, the branes and antibranes do not
annihilate because of the finite temperature of the gas which gives a positive mass squared to the open string tachyon. It can be shown that for $T>1/\sqrt{g_s N} \ell_s$ the system
is stable. In addition, one needs to assume that $T< \ell_s^{-1}$ so that excited open strings can be neglected justifying the use of field theory on the brane world--volume.
Due to the finite temperature, the scalars and the fermions in the field theory get masses $O(\sqrt{g_s N} T)$ and are decoupled. The only remaining massless degrees of freedom are
$O(N^2)$ gauge bosons. In addition, the two gases are decoupled because strings that connect the branes to antibranes (the tachyon which is stabilized) cannot be excited at these low 
temperatures. The mass, entropy and charge of the $D3$--${\bar D}3$ system is given by
$$M_{FT}=(N+\bar N) T_3 V+a{\pi^2 \over 16} N^2 T^4 V+a{\pi^2 \over 16} {\bar N}^2 T^4 V \eqno(2.1)$$
$$S_{FT}=a{\pi^2 \over 12} N^2 T^3 V+a{\pi^2 \over 12} {\bar N}^2 T^3 V \eqno(2.2)$$
and
$$Q=N- \bar N \eqno(2.3)$$
where the tension of the $D3$ brane is $T_3=1/(2 \pi)^3 g_s \ell_s^4$. The constant $a=8$ in weakly coupled super Yang--Mills theory but can be argued to be $a=6$ in the strongly coupled 
limit from AdS/CFT correspondence.

We now specialize to the \S black hole which is described by $N=\bar N$. Then the gases on the branes and antibranes have the same energy and temperature. Eqs. (2.1) and (2.2) become
$$M_{FT}=2N T_3 V+ a{\pi^2 \over 8} N^2 T^4 V \eqno(2.4)$$
and
$$S_{FT}=a {\pi^2 \over 6} N^2 T^4 V \eqno(2.5)$$
The value of $N$ is obtained by maximizing the entropy for fixed $M_{FT}$. Maximizing the entropy
$$S_{FT}=a^{1/4}2^{5/4}3^{-1} \sqrt {\pi N} V^{-1/4}(M_{FT}-2N T_3 V)^{3/4} \eqno(2.6)$$
one finds that
$$N={1 \over 5} {M_{FT} \over {T_3 V}} \eqno(2.7)$$
We see that the energy in the brane tension and in the gas are comparable. Using eq. (2.7) in eq. (2.6) we find
$$S_{FT}=a^{1/4} 2^{5/4} 3^{-1/4} 5^{-5/4} \pi^{1/4} \sqrt \kappa V^{-1/4} M_{FT}^{5/4} \eqno(2.8)$$
where $T_3=\sqrt \pi/\kappa$ and $G_{10}=8 \pi/\kappa^2$. From supergravity, a neutral black $D3$ brane with \S radius $r_0$ has mass
$$M_{SG}={5 \over 2} {\pi^3 \over \kappa^2} r_0^4 V \eqno(2.9)$$
Then, using $M_{FT}=M_{SG}$ we get
$$S_{FT}={{2^{1/4} \pi^4} \over \kappa^2} r_0^5 V=2^{-3/4}S_{SG} \eqno(2.10)$$
The temperature of the gas, $T_{FT} \sim (g_s N)^{-1/4} \ell_s^{-1}$ gives the Hawking temperature which satisfies $1/\sqrt{g_s N} \ell_s^{-1} <<T<<\ell_s^{-1}$ as required. 
The temperature of the gas is not
crucial in this case because the world--volume theory on a $D3$ brane (or antibrane) is \s at all values of the coupling. We will see in the next section that the low gas temperature
requires the world--volume theory to be the \s theory rather than the naive $p$ dimensional ideal gas. 
The negative specific heat of the black hole is explained by the fact that when the energy decreases, brane--antibrane pairs annihilate. The energy of the pair goes to the gas which increases
its temperature resulting in negative specific heat for the system. 


The above description can be generalized to $D=7$ charged black holes (or to black $D3$ branes wrapped on $T^3$)[\ULF]. The mass, entropy and charge of the system is given by eqs. 
(2.1)-(2.3) with $N \not= \bar N$. The supergravity expressions for these are given by
$$M_{SG}={\pi^3 \over \kappa^2} r_0^4 V (cosh 2 \gamma+{3 \over 2}) \eqno(2.11)$$
$$S_{SG}={{2 \pi^4} \over \kappa^2} r_0^5 V cosh \gamma \eqno(2.12)$$
$$Q={\pi^{5/2} \over \kappa} r_0^4 sinh 2 \gamma \eqno(2.13)$$
Comparing these to eqs. (2.1)-(2.3) we find the number of branes and antibranes
$$N={\pi^{5/2} \over{2 \kappa}} r_0^4 e^{2 \gamma} \qquad \qquad {\bar N}={\pi^{5/2} \over{2 \kappa}} r_0^4 e^{-2 \gamma} \eqno(2.14)$$
and the gas energy
$$E={3 \over 2}{\pi^3 \over \kappa^2} r_0^4 V \eqno(2.15)$$
In terms of $N, \bar N$ and $E$ the entropy becomes
$$S_{SG}=2^{5/4} 3^{-3/4} \pi^{1/2} V^{1/4} E^{3/4}(\sqrt N+ \sqrt{\bar N}) \eqno(2.16)$$
which is the entropy of a gas on $N$ $D3$ branes plus that of a gas on $\bar N$ ${\bar D}3$ branes. Note that for $N \not= \bar N$, the temperatures of the two gases are not equal. This
is not a problem because the two gases are decoupled at these low temperatures. In this case, $T_{FT}$ is given by
$${2 \over T_{FT}}={1 \over T}+{1 \over {\bar T}} \eqno(2.17)$$
and agrees with the Hawking temperature.
A point that is left unexplained is the fact that for the prescription to work the two gases need to have equal energies even though they are not in contact[\ULF].

\bigskip
\centerline{\bf 3. \S Black Hole Entropy and Superconformal Field Theories}
\medskip

In this section we generalize the results of ref. [\ULF] summarized in the previous section to \S black holes in $D>4$. As we will see, these black holes can be described by 
brane--antibrane systems of $p=10-D$ dimensions with a superconformal gas (at a finite temperature) living on them. Again the only massles degrees of freedom arise from strings that 
connect branes to branes or antibranes to antibranes. We assume that, in all dimensions, the temperature is high enough to 
stabilize the system, i.e. to give a positive mass squared to the open string tachyon of the system. As a result, the strings that connect branes to antibranes are massive and the branes 
are decoupled from the antibranes. The reason the gas is \s and not an ideal gas is its low temperature; 
in all cases the dimensionless ('t Hooft) coupling of the world--volume theory is very large. This means that the world--volume field theory is described by a \s field theory rather than 
the weakly coupled super Yang--Mills theory. 

We first consider the specific cases of \S black holes in $D=9,6,5$ dimensions in detail. These are described by \s gases on brane--antibrane systems with $p=1,4,5$. Since, in the strong
coupling limit, we know the equation of state of the \s theories up to numerical factors of $O(1)$, all our results for black hole entropy will also be correct up to the same factors. However,
if one uses the AdS/CFT duality to derive the numerical coefficients of the equations of state the mismatch between $S_{FT}$ and $S_{SG}$ becomes a computable power of $2$[\PEE,\GIL,\KRS]. 

{\bf 3.1. $D=9$ \S Black Holes}

We begin with the \S black hole in $D=9$. According to the prescription of ref. [\ULF], this can be described by two copies of a gas that lives on $N$ $D1$ branes and $\bar N=N$ 
anti--$D1$ branes. In $1+1$ dimensions, the coupling constant has dimensions of mass. Therefore the dimensionless ('t Hooft) coupling constant of the gas is
$$\tilde g^2={{g^2 N} \over T^2} \eqno(3.1)$$
We will see below that $\tilde g^2 \sim N^{4/3}$. At strong coupling, the gas is described by a \s theory with equation of state[\KLE,\KLS]
$$S \sim N^{3/2} L T^2 g^{-1} \qquad \qquad E \sim N^{3/2} L T^3 g^{-1} \eqno(3.2)$$
or using $e=g \sqrt N$
$$S \sim N^2 L T^2 e^{-1} \qquad \qquad E \sim N^2 L T^3 e^{-1} \eqno(3.3)$$
where the $D1$ brane tension is $T_1=1/(2 \pi) g_s \ell_s^2$.
We see that the above equations of state are those of a $2+1$ dimensional \s theory with a new dimension of size $g^{-1}$. This is not surprising since $D1$ branes are in fact $M2$ branes
(wrapped around a shrinking $T^2$). In fact, the equation of state in eq.(3.3) is exactly the one for an ideal gas for $M2$ branes (considered in ref. [\ULF]) 
with a volume of $L G^{-1}$. Thus $D=9$
\S black holes can be described either by a system of $M2-\bar M2$ branes in M--theory or a system of $D1-\bar D1$ branes in string theory.
Now we can write the mass of the $N$ pairs of $D1$ branes and antibranes and the two gases living on them as
$$M_{FT}=2N L T_1+2N^2 LT^3 e^{-1} \eqno(3.4)$$
The entropy is
$$S_{FT}= 2N^2 L T^2 e^{-1} \eqno(3.5)$$
Maximizing the entropy with respect to $N$ 
$$S=(M_{FT}-2NLT_1)^{2/3} 2^{4/3}N^{1/2} L^{1/3} g^{-1/3} \eqno(3.6)$$
for a given $M_{FT}$ we find that
$$N={3 \over 5 }{M_{FT} \over {LT_1}} \eqno(3.7)$$
Using this in eq. () and taking $M_{FT}=M_{BH}$ (with $G_9=(g_s^2 \ell_s^8/L)^{-1}$)  we find 
$$S_{BH} \sim M_{BH}^{7/6} G_9^{1/6} \eqno(3.8)$$
which is the entropy of a $D=9$ \S black hole.
Alternatively using the supergravity mass for the black hole with \S radius $r_0$[\HST,\KLE]
$$M_{BH}={{7 \omega_7} \over {2 \kappa^2}} L r_0^6 \eqno(3.9)$$
we find that
$$S_{BH} \sim {\omega_7 \over {2 \kappa^2}} L r_0^7 \eqno(3.10)$$ 
Using eqs. (3.6) and (3.7) we find the temperature of the gas (which is the Hawking temperature) 
$$T=({2 \over 3} N L T_1)^{1/3} N^{-2/3} L^{-1/3} e^{1/3} \sim {{g^{1/3} N^{-1/6}} \over \ell_s^{2/3}} \eqno(3.11)$$
This temperature used in eq. (3.1) justifies our argument for a \s theory for the gas.

{\bf 3.2. $D=6$ \S Black Holes}

For a $D=6$ \S black hole we need to consider two copies of \s gas on $N$ pairs of $D4$--$\bar D4$ branes. Thus we need to describe a $4+1$ dimensional \s field theory in which the coupling
has dimension (length)$^{1/2}$. Such a theory has an equation of state given by[\KLE,\KLS]
$$S \sim N^3 L^4 T^5 g^2 \qquad \qquad E \sim N^3 L^4 T^6 g^2 \eqno(3.12)$$
or using $e=g \sqrt N$
$$S \sim N^2 L^4 T^5 e^2 \qquad \qquad E \sim N^2 L^4 T^6 e^2 \eqno(3.13)$$
where the $D4$ brane tension is $T_4=1/(2 \pi)^4 g_s \ell_s^5$.
We see that the equation of state describes a \s theory in $5+1$ dimensions with an extra dimension of size $g^2$[\ROZ,\BRS]. Again, this is not surprising since a $D4$ brane is an $M5$ brane
wrapped around the $11^{th}$ dimension. The equation of state in eq. (3.13) is exactly the one for an ideal gas (considered in ref. [\ULF]) on $M5$ branes with volume $L^4 g^2$. Therefore we
can describe $D=6$ \S black holes either by a system of $M5-\bar M5$ branes in M--theory or by a system of $D4-\bar D4$ branes in string theory.
The mass of the brane--antibrane system is
$$M_{FT}=2N L^4 T_4+2N^2 L^4 T^6 e^2 \eqno(3.14)$$
whereas the entropy is
$$S \sim 2N^2 L^4 T^5 e^2 \eqno(3.15)$$
Eliminating the temperature from $S$ we find
$$S \sim (M-2N L T_4)^{5/6} 2^{7/6}N^{1/3} L^{2/3} e^{1/3} \eqno(3.16)$$
Maximizing $S$ with repect to $N$ for a fixed $M_{FT}$ we find
$$N={6 \over 11}{M_{FT} \over {L^4 T_4}} \eqno(3.17)$$
Using the expression for $N$ in $S$ and taking $M_{FT}=M_{BH}$ we get
$$S_{BH} \sim M_{BH}^{4/3} G_6^{1/3} \eqno(3.18)$$
where $G_6=g_s^2 \ell_s^8/L^4$. Alternatively, we can use the supergravity mass of the black hole with \S radius $r_0$[\HST,\KLE]
$$M_{BH}={{4 \omega_4} \over {2 \kappa^2}} L^4 r_0^3 \eqno(3.19)$$
we find that
$$S_{BH} \sim {\omega_4 \over {2 \kappa^2}} L^4 r_0^4 \eqno(3.20)$$ 
which is the correct result. In this case the temperature of the gas is
$$T =({5 \over 6} N L^4 T_4)^{1/6} N^{-1/3} L^{-2/3} e^{-1/3} \sim N^{-1/6} \ell_s^{-5/6} g^{-1/3}  \eqno(3.21)$$
As a result, the dimensionless ('t Hooft) coupling is very large
$$\tilde g^2 = g^2 N T \sim g^{5/3} N^{2/3} \ell_s^{-5/6} \eqno(3.22)$$
which justifies our use of the \s theory.

{\bf 3.3. $D=5$ \S Black Holes}

For $D=5$ \S black holes we need to consider two copies of gas on $N$ pairs of $D5$--$\bar D5$ branes. It is well--known the world--volume theory on $D5$ branes is not a field theory
but a (noncritical) string theory[\MST]. In fact, the finite temperature gas on $N$ $D5$ branes can be described by $N$ strings with fractional tension $\sim 1/g^2 N \sim 1/e^2$ at their
Hagedorn temperature $T_H \sim 1/g \sqrt N \sim 1/e$. As a result we have (with the $D5$ brane tension $T_5=1/(2 \pi)^5 g_s \ell_s^6$) 
$$M_{FT}=2N T_5 L^5 + 2N {1 \over {g \sqrt N}} \eqno(3.23)$$
for the mass and 
$$S=N=g \sqrt N(M_{FT}-2N T_5 L^5) \eqno(3.24)$$
Maximizing $S$ we find
$$N={1 \over 3}{M_{FT} \over {T_5 L^5}} \eqno(3.25)$$
so that now
$$S={2 \over 3} g \sqrt N M \eqno(3.26)$$
We see that entropy is proportional to energy as expected in a string theory. The Hagedorn temperature is $T_H \sim 1/g \sqrt N$.
Using eq. (3.25) and $M_{FT}=M_{BH}$, $S$ can be written as 
$$S \sim M_{BH}^{3/2} G_5^{1/2} \eqno(3.27)$$
where $G_5=(g_s^2 \ell_s^8/L^5)^{-1}$.
Alternatively, we can take the supergravity mass of of the \S black hole with radius $r_0$[\HST,\KLE]
$$M_{BH} \sim {{3 \omega_3} \over {2 \kappa^2}} L^5 r_0^2 \eqno(3.28)$$
we find that
$$S_{BH} \sim {\omega_3 \over {2 \kappa^2}} L^5 r_0^3 \eqno(3.29)$$ 
as required. Note that in this case the dimensionless coupling is
$$\tilde g^2= g^2 N T^2 \sim 1 \eqno(3.30)$$

{\bf 3.4 \S Black Holes in $D>4$}

From the above discussion it is clear that, in general, a $D>4$--dimensional \S black hole can be described by a system of $N$ $p=10-D$ brane--antibranes with a \s gas on their
world--volumes. The generic equation of state for a \s theory on $p+1=11-D$ dimensions is[\KLE,\KLS]
$$S \sim \sqrt N E^{\lambda} L^{(10-D)(1-\lambda)} g^a \eqno(3.31)$$
where
$$2 \lambda={{D-1} \over {D-3}} \qquad \qquad a={{7-D} \over {D-3}} \eqno(3.32)$$ 
and $E$ is the energy of the \s gas $E=M_{FT}-N T_p L^p$.
As we saw above, in all dimensions, maximizing the entropy gives
$$N \sim {M_{FT} \over {T_p L^p}} \eqno(3.33)$$
Then, the entropy becomes (with $M_{FT}=M_{BH}$)
$$S \sim M_{BH}^{(D-2)/(D-3)} G_D^{1/(D-3)} \eqno(3.34)$$
which is the correct entropy for \S black holes in $D>4$ dimensions.
We can also use the supergravity mass of the \S black hole with radius $r_0$[\HST,\KLE]
$$M_{BH} \sim {\omega_{D-2} \over {2 \kappa^2}} L^p r_0^{D-3} \eqno(3.35)$$
to obtain
$$S \sim {\omega_{D-2} \over {2 \kappa^2}}L^p r_0^{D-2} \eqno(3.36)$$
The temperature of the gas is obtained from eq. (3.31) to be
$$T^{-1} \sim E^{\lambda-1/2} L^{(10-D)(1-\lambda)-p/2} g^a T_p^{-1/2} \eqno(3.37)$$
It is easy to show that $T$ is in fact the Hawking temperature of the black hole. For a large black hole $T \sim E^{1/2-\lambda} \sim N^{1/2-\lambda}$ is very low which justifies our use
of the \s theories. 
We also see that $\lambda>1/2$ for all $D$ and therefore $T$ and $E$ are inversely proportional resulting in a negative specific heat for the black holes.

We now show that the \S radius of the black hole (for any $D>4$) is given by the transverse oscillations of the branes (and antibranes) due to the finite temperature. From the virial
theorem we get for the branes
$$N m_p <\dot{X}^2> \sim g^{(p-3)/(5-p)} T^{(14-2p)/(5-p)} L^p N^{(7-p)/(5-p)} \eqno(3.38)$$
where the mass of a $p$--brane is $m_p=L^p/(2 \pi)^p \ell_s^{p+1}$. Now using the fact that the characteristic frequency of the branes is $\sim T$, we can replace the time derivatives with
$T$. The maximization of the entropy of the system with respect to $N$ gives
$$N \sim {m_{FT} \over {L^p T_p}} \eqno(3.39)$$
Assuming $M_{FT}=m_{SG}$ and using $T \sim 1/r_0$ and $m_{SG} \sim r_0^{8-p}/G_{10-p}$ (with $G_{10-p}=g_s^2 \ell_s^8/L^p$) we find from (3.38) and (3.39) that $<X^2> \sim r_0^2$. 
Since the temperature of
the antibranes is equal to that of the branes we get the same result for the antibranes. Therefore, the extent of the transverse oscillations of the brane--antibrane system due to the
finite temperature gives black hole radius. For $D=5$ (with $p=5$), eq. (3.38) does not make sense. However, in this case the thermal fluctuations of the branes are given by those
of the fractional strings. For a fractional string with tension $\sim 1/g^2 N$ and energy $\sim T \sim 1/g \sqrt N$, the extent of the transverse fluctuations is $g \sqrt N \sim r_0$.

Our results for the black hole entropy hold up to numerical factors of $O(1)$. This is due to our lack of knowledge of the precise equation of state for the \s theories in
different dimensions. Since these theories are strongly coupled and supersymmetry is broken by the finite temperature of the gas, the equations of state may get corrections of $O(1)$.
If we assume that the equation of state for the \s theories are given exactly (up to the numerical factor) by near extreme black branes, we find that $S_{FT}=2^{-(9-p)/2(8-p)} S_{SG}$
which is a small mismatch. This assumption is well--motivated by the AdS/CFT duality; however it is not clear that the numerical coefficients derived from it are correct in the region
of the parameter space we are interested in. 

The entropy of \S black holes in $D=4$ cannot be obtained by the prescription of ref. [\ULF] because $D6$ branes have world--volume theories with negative specific heat. As a result, 
world--volume theory does not decouple from the bulk. 

\bigskip
\centerline{\bf 4. Charged Black Hole Entropy and Superconformal Field Theories}
\medskip

We now generalize our results to charged black holes or (wrapped) black branes. For a $D$--dimensional charged black hole we need to consider a system of $N$ $Dp$ branes together with $\bar N$
anti $Dp$ branes where $N \not =\bar N$[\ULF]. In addition there will be two \s gases at different temperatures living on the branes and antibranes. As for the $D=7$ case the two gases will have
different temperatures but need to have the same energy in order to reproduce the correct entropy.
The mass, entropy and charge of the $Dp$--${\bar D}p$ system is given by
$$M_{FT}=(N+\bar N) T_p L^p + g^{(p-3)/(5-p)} T^{(9-p)/(5-p)} L^p (N^{(7-p)/(5-p)}+ {\bar N}^{(7-p)/(5-p)}) \eqno(4.1)$$
$$S_{FT}=g^{(p-3)/(5-p)} T^{4/(5-p)} L^p (N^{((7-p)/(5-p)}+ {\bar N}^{((7-p)/(5-p)}) \eqno(4.2)$$
and
$$Q=N- \bar N \eqno(4.3)$$
where the tension of the $Dp$ brane is $T_p=1/(2 \pi)^p g_s \ell_s^{p+1}$. 
Again, both gases are at very low temperatures which mean that the dimensionless 't Hooft couplings are
very large. This justifies our use of the \s theories on the brane world--volumes.
The equations of state for each one of the \s gases is 
$$S \sim \sqrt N E^{\lambda} L^{(10-D)(1-\lambda)} g^a \eqno(4.4)$$
where $E$ is the energy of the \s gas $2E=M_{FT}-N T_p L^p-\bar N T_p L^p$.
The sum of the entropies of the two gases is given by
$$S \sim (\sqrt N + \sqrt {\bar N})  E^{\lambda} L^{(10-D)(1-\lambda)} g^a \eqno(4.5)$$
where we used the fact that both gases have the same energy. The supergravity mass, entropy and charge of a $D$--dimensional black hole are given by ($d=7-p$)[\HST,\KLE]
$$M_{SG}={\omega_{d+1} \over {2 \kappa^2}} r_0^d L^p (sinh^2 \gamma+d+1) \eqno(4.6)$$
$$S_{SG}={\omega_{d+1} \over {2 \kappa^2}} r_0^{d+1} L^p cosh \gamma \eqno(4.7)$$
$$Q={\omega_{d+1} \over {4 \kappa^2}} r_0^d L^p sinh 2 \gamma \eqno(4.8)$$
As before, the relations between $N,\bar N$ and $E$ are obtained by maximizing the entropy with respect to $N$ and $\bar N$.
$$E={{2 (D-1)} \over (D-3)} \sqrt{N \bar N} \eqno(4.9)$$
From the expression for the charge we find that
$$N={\omega_{d+1} \over {4 \kappa^2}} r_0^d L^p e^{2 \gamma} \qquad \qquad \bar N={\omega_{d+1} \over {4 \kappa^2}} r_0^d L^p e^{-2 \gamma} \eqno(4.10)$$
Using the expressions for $E$ an $N, \bar N$, we find that the entropy of the \s gas in eq. (4.4) agrees with that of the charged black hole in eq. (4.7). If we use the results from 
AdS/CFT duality to predict the coefficients
of the equations of state for the \s theories we find $S_{FT} 2^{\lambda} S_{SG}$ where is given by eq. (3.32).

In this case the temperature $T$, on the branes and $\bar T$ on the antibranes are not equal; it can be shown that $T \sim 1/r_0 e^{\gamma}$ and $\bar T \sim 1/r_0 e^{- \gamma}$. For $D=5$
black holes, we have two types of fractional strings on branes and antibranes with two different tensions, $1/g^2N$ and $1/^2 \bar N$ leading to two different Hagedorn temperatures.
The temperature of the system given by eq. (2.17) agrees with the Hawking temperature as expected. Note that as the black hole approaches the near extreme limit, the temperature of the
gas on the antibranes increases. This is due to the fact that as we get closer to extremality, branes and antibranes annihilate and decrease the value of $N$ and $T \sim N^{-1/(D-3)}$.
Obviously, for near extreme black holes (with $E< T_p L^p$) the above results should agree with those in the literature.

Again the \S radius of the black hole is given by the transverse thermal fluctuations of the branes and antibranes. Note that now $T \not = \bar T$ and therefore we need to consider the 
branes and antibranes separately. Let us consider the branes first. Eq. (3.38) for the brane energy remains the same. However, now eq. (3.39) for the number of branes is modified to 
$$N \sim {m_{FT} \over {L^p T_p}} e^{2 \gamma} \eqno(4.11)$$
In addition, now $T \sim 1/r_0 e^{\gamma}$. Using the relation between $M_{SG}$ and $r_0$ (and those for $m_p$, $G_{10-p}$) we find $<X^2> \sim r_0^2$ as before. For the antibranes,
we obtain $,\bar X^2> \sim r_0^2$ if we repeat the same exercise but switch the sign of $\gamma$ in eq. (4.11) and use $\bar T \sim 1/ r_0 e^{- \gamma}$.

\bigskip
\centerline{\bf 5. Multicharged Black Hole Entropy and Superconformal Field Theories}
\medskip

For $D$--dimensional black holes with $n$ charges the mass, entropy and charge are given by[\HST,\KLE]
$$M_{SG}={\omega_{d+1} \over {2 \kappa^2}}d r_0^d L^p (\sum_{i=1}^n cosh \gamma_i+2\lambda) \eqno(5.1)$$
$$S_{SG}={\omega_{d+1} \over {2 \kappa^2}} r_0^{d+1} L^p \prod_{i+1}^n cosh \gamma_i \eqno(5.2)$$
$$Q_i=N_i-\bar N_i ={\omega_{d+1} \over {4 \kappa^2}} r_0^d L^p sinh 2 \gamma_i \eqno(5.3)$$
where
$$2 \lambda={{D-1} \over {D-3}}-{n \over 2} \eqno(5.4)$$ 

Comparing eqs. (5.1)--(5.3) with (4.1) and (4.3) for one charge, we see that the entropy of a black hole with $n$ charges can be described by a collection of $n$ types of 
brane--antibrane systems
with different \s gases on each pair of $p p^{\prime}$ branes and/or antibranes. In this picture, basically, every $p p^{\prime}$, $\bar p p^{\prime}$, $p \bar p^{\prime}$ and 
$\bar p \bar p^{\prime}$ type of string describes a different \s gas. Note that the $p \bar p$ strings which constituted the \s gases in neutral and one--charged black holes 
do not form \s gases and do not contribute to the entropy.
For $n$ charges, there are $2^n$ \s gases arising from the $2^{n-1}$ brane pairs with a pair of \s gases on each one. As before, we assume that this
collection of branes and antibranes is stable, i.e. there are no open string tachyons due to the finite temperature.

Even though the above counting of \s gases works it is difficult to understand it from D--brane physics. For neutral and one--charged black holes the \s gases arose from strings connecting
the branes to branes and antibranes antibranes. These gauge bosons are clearly massless even at finite temperature. The world--volume fermions and scalars and the strings connecting the 
branes to the antibranes are massive due to the finite temperature. In order to see the problem with the current case, consider a $D=5$ black hole with two charges. This can be obtained
from a system of $N_5$ ($\bar N_5$) D5 branes (anti--D5 branes) and $N_1$ ($\bar N_1$) D1--branes (anti--D1 branes) wrapped on $T^5$. Now, since the D1 branes are inside the D5 branes,
the system will be in the Higgs branch (at zero temperature) and therefore the gauge bosons are massive. In addition the world--volume fermions and scalars arising from $p \bar p$ strings are
massive due to the finite temperature. At zero $T$, the system would be described by the moduli of the Higgs branch which are given by all the massless modes arising from $51$,$15$, $\bar 5 1$
and $5 \bar 1$ strings. However, these are world--volume scalars and fermions and should be massive at finite $T$. How they remain massless and form \s gases is unfortunately not clear.

Each one of the \s gases has an equation of state
$$S \sim \sqrt N E^{\lambda} L^{(10-D)(1-\lambda)} g^a \eqno(5.5)$$
where $E$ is the energy of the \s gas.
From eq. (5.3) we find that
$$N_i={\omega_{d+1} \over {4 \kappa^2}} r_0^d L^p e^{2 \gamma_i} \qquad \qquad \bar N_i={\omega_{d+1} \over {4 \kappa^2}} r_0^d L^p e^{-2 \gamma_i} \eqno(5.6)$$
Then the total entropy of the system becomes
$$S \sim \prod_{i=1}^n(\sqrt N_i + \sqrt {\bar N_i})  E^{\lambda} L^{(10-D)(1-\lambda)} g^a \eqno(5.7)$$
where we took into account all $2^n$ \s gases. As usual, maximizing the entropy for a given mass and set of charges we find
$$E={{2 (D-1)} \over (D-3)}  [\prod_{i=1}^n(N_i \bar N_i)]^{1/2} \eqno(5.8)$$
Using eqs. (5.6) and (5.8) we find that the total entropy of the gas given by eq. (5.7) equals (up to a numerical constant) that of the black hole in supergravity. Again, if we use the AdS/CFT 
duality to predict the numerical coefficient in the equations of state, we find $S_{FT}=2^{-\lambda n} S_{SG}$ where $\lambda$ is given by eq. (3.32).
Note that the total entropy is the sum of the entropies of the $2^n$ \s gases which means that the \s gases do not interact with each other.
The temperature of the collection $T_{FT}$ which is the Hawking temperature of the black hole is given by
$${n \over T_{FT}}=\sum_{j=1}^{2^n} {1 \over T_j} \eqno(5.9)$$
where $2 \leq j \leq 2^n$ counts all the possible combinations of branes and/or antibranes.

\bigskip
\centerline{\bf 6. Conclusions and Discussion}
\medskip

In this paper we showed that the entropy of \S and charged black holes in $D>4$ can be obtained from that of a system of $N$ branes and $\bar N$ antibranes with a pair of \s gases living on 
them. This is a 
generalization of the prescription of ref. [\ULF] to different dimensions. Note that in [\ULF], the simplest case with $D=7$ or $p=3$ which is \s for any value of $g$ was examined.
Since we do not 
know the precise equation of state of the \s field theory in $p<6$ dimensions, our results give the black hole entropy up to numerical factors of $O(1)$. In all dimensions, the 
world--volume theory is in the strongly coupled \s limit (rather than say in the weakly coupled super Yang--Mills limit) due to the low temperature. In other words, in all cases,
the dimensionless 't Hooft coupling at the scale given by the temperature of the gas is very large. Well--known but nevertheless intriguing properties of these \s theories are crucial
for reproducing the black hole entropy. These include the appearance of new dimensions for $p=1,4$ and of fractional strings for $p=5$. Not surprisingly, we were not able to obtain the 
entropy of $D=4$ black holes due to the well--known fact that D6 brane world--volume theory has a negative specific heat and is not well--defined. We also showed that the same prescription
can explain the entropy of multicharged black holes. For a black hole with $n$ charges, there are $2^n$ \s gases of the types described above which do not interact with each other.

As shown in refs. [\GUI.\PEE], one can also add angular momentum to the black holes considered above. in this case, on the world--volume, one needs to consider \s gases which carry R--charges. 
The nonzero R--charge requires a modification of the equations of state of the \s theories which can be obtained from the AdS/CFT duality.

In ref. [\GIL] it was noticed that the entropy of black holes far from extremality can be obtained from the entropy of near extreme D--branes using the prescription of ref. [\ULF]. 
In light of our
results this makes sense since the equations of state for \s theories that we used can be obtained from a correspondence between near extreme branes and world--volume theories at a finite
temperature. The same equations of state have been used to obtain the entropy of $D>5$ \S black holes in M(atrix) theory[\KLS]. In that case the world--volume theory is near extreme due 
to the large 
boost. In the present case, it seems that a black hole far from extremality can be constructed by adding a large number of near extreme brane--antibrane pairs with zero net charge.

Even though the above set up is very different than that in M(atrix) theory, the superficial similarities are intriguing. For example, M(atrix) theory gives the entropy of $D$--dimensional
\S black holes in terms of a $p=11-D$--dimensional \s gas[\MAT-\DAS] whereas we saw above that the same gas describes a \S black hole in $D=10-p$-dimensions. 
This difference of one dimension is not 
surprising since we are working in string theory but M(atrix) theory describes $D=11$ M--theory. Note, however, that the temperature of the gases and the relation between $S$ and $N$ are 
different in the two cases. It would be interesting to see if these two description of \S black holes are related.

The prescription of ref. [\ULF] depends on the fact that brane--antibrane pairs are stable at a finite temperature. This was shown for $p=3$ in detail in [\ULF]. In this paper, we have
simply assumed that this is still the case for $p<6$ in order to use the same prescription. It would be nice if this can be shown explicitly by examining the behavior of the open string 
tachyon at a finite temperature for all $p$. 

Finally, due to the success of the above prescription in explaining black hole entropy one may wonder whether a similar approach may be used to explain the entropy of de Sitter space.
In this respect, the fact that de Sitter space entropy can be explained by long strings on the stretched horizon just like \S black holes seems encouraging[\DES].

\bigskip
\centerline{\bf Acknowledgements}

I would like to thank Lenny Susskind for a very useful discussion.

\vfill

\refout

\end
\bye